\documentclass[12pt]{article}

\usepackage[T1]{fontenc}
\usepackage[utf8]{inputenc}
\usepackage{amsmath, amssymb}
\usepackage{microtype}          
\usepackage{xurl}               
\usepackage[hidelinks]{hyperref} 
\usepackage{longtable}

\usepackage[letterpaper,margin=1in]{geometry} 
\usepackage{setspace}
\usepackage{booktabs,threeparttable,siunitx,array}
\usepackage{float}    
\usepackage{placeins} 
\usepackage{graphicx}
\usepackage{appendix}

\usepackage{titlesec}
\titlespacing*{\section}{0pt}{1.2\baselineskip}{0.6\baselineskip}
\titlespacing*{\subsection}{0pt}{1.0\baselineskip}{0.4\baselineskip}
\titlespacing*{\subsubsection}{0pt}{0.8\baselineskip}{0.3\baselineskip}

\title{Pixels to Prices: Visual Traits, Market Cycles, and the Economics of NFT Valuation}
\author{Samiha Tariq\\
\small PhD Candidate, Economics\\
\small Southern Illinois University Carbondale}
\date{\today}

\begin{document}
\maketitle

\begin{abstract}
  Pixels and market cycles both move NFT prices. Non-fungible tokens (NFTs) are unique digital assets, often used to represent ownership of digital art, collectibles, and other media, secured on blockchain networks like Ethereum. The rise of NFTs has led to the creation of a multi-billion-dollar market for digital art and collectibles, making it a key area of interest for researchers, artists, and investors. Using 94,039 transactions from 26 major generative Ethereum collections, this study extracts 196 machine-quantified image descriptors—color, composition, palette structure, geometry, texture, and deep-learning embeddings—and applies a three-stage filter to identify stable predictors for hedonic regression. A static mixed-effects model shows that market sentiment and transparent, interpretable image traits have significant and independent pricing power: higher focal saturation, tighter compositional concentration, and greater curvature are rewarded, while clutter, heavy line work, and dispersed palettes are discounted; deep embeddings add limited incremental value once explicit traits are included. To assess state dependence, a Bayesian dynamic mixed-effects panel with cycle effects is estimated, allowing Composition Focus—Saturation—the ratio of saturation in the central region to the whole image, capturing vividness and concentration at the focal area—to vary across market regimes. Collection-level heterogeneity (brand premia) is absorbed by random effects. The time-varying coefficients exhibit clear regime sensitivity, with stronger premia in expansionary phases and weaker or negative loadings in downturns, while the grand-mean effect is small on average. Overall, NFT prices reflect both observable digital product characteristics and market regimes, and the framework offers a cycle-aware tool for asset pricing, platform strategy, and market design in digital art markets.
  \end{abstract}

\section{Introduction}

Digital assets are traded, valued, and analyzed at a scale once reserved for conventional art and finance. Nowhere is this clearer than in the explosive growth of NFTs—non-fungible tokens recorded as unique entries on public blockchains via standards such as Ethereum’s ERC-721, which define how singular, non-interchangeable tokens are created, owned, and transferred (EIP-721, 2018; Kaisto, 2024). The rapid rise of generative NFT collections has created a new empirical laboratory: thousands of algorithmically produced artworks, sold in transparent secondary markets, with every transaction and visual trait recorded in public ledgers.

Economists and market analysts have asked: what determines prices in this new domain? Classic art markets reward visible traits—composition, color, clarity, signature—but digital assets are traded with near-complete information, at high frequency, and within evolving speculative cycles. The challenge is not only to identify which features matter, but to separate persistent trait effects from the strong ebbs and flows of sentiment that characterize crypto markets. Empirical work documents both substantial pricing premia linked to attributes and episodes of inefficiency typical of nascent markets (Dowling, 2022; Alsultan, Kourtis, \& Markellos, 2024; Ante, 2022).

Existing studies in art and asset pricing emphasize hedonic models, in which product attributes and environmental factors jointly predict value. In the context of NFTs, this approach collides with two facts. First, image features are quantifiable at scale: modern computer vision can extract hundreds of attributes per artwork, beyond “rarity” tags (Alsultan et al., 2024). Second, NFT markets are highly dynamic, with repeated booms and busts; comovement with crypto benchmarks is documented but incomplete, and spillovers vary across episodes (Dowling, 2022; Ante, 2022).

The market itself is economically meaningful. On Ethereum’s dominant marketplace, OpenSea, cumulative traded volume exceeds \textbf{\$40 billion} in USD terms as of September 2025, underscoring the scale at which these assets are exchanged (Dune Analytics, OpenSea dashboard). Monthly marketplace volume on Ethereum remains material even after incentive changes and cyclical slowdowns, with industry trackers reporting persistent multi-billion-dollar turnover filtered for wash trading (The Block, Marketplace Monthly Volume). These magnitudes make NFTs a suitable setting to study how observable traits and market regimes jointly shape price discovery.

Prior research has explored rarity, collection-level traits, and broad price patterns in NFTs, but most analyses stop short of integrating quantified image features and explicit market cycles. Econometric models that allow for time-varying coefficients—especially alongside high-dimensional traits—remain rare in this literature, limiting both theory and practice for artists, investors, and platforms.

This paper addresses both challenges directly. The analysis builds a dataset of 94{,}039 secondary-market sales from 26 major generative Ethereum collections, characterizing each asset by 196 machine-quantified visual features: color metrics, palette structure, compositional clarity, line geometry, texture statistics, and deep-learning embeddings. The large sample allows clean separation of trait effects from market noise and unobserved collection heterogeneity. To link pixels to prices, the study uses a three-stage selection process to identify robust predictors for mixed-effects hedonic regressions, distinguishing persistent, interpretable trait premia from background shocks. The paper then introduces a Bayesian dynamic panel with cycle effects and a single time-varying coefficient for a salient image attribute, testing whether "what sells" is a stable law or a moving target shaped by sentiment and liquidity.

The findings show that both image features and market cycles matter—and their effects interact. NFTs with higher focal saturation, clean composition, and smooth curvature command consistent premia, even after conditioning on collection and market regime. Yet trait premia are not fixed: they strengthen in expansionary periods and compress or reverse in downturns. Deep-learning embeddings add little once explicit traits are included, suggesting that transparent, interpretable features dominate in this market. The results bridge three literatures—art valuation, empirical asset pricing, and digital economics—by showing that NFT prices reflect recognizable mechanisms: observable attributes and time-varying sentiment. In short, both pixels and cycles shape price discovery; the right model sees both.

\section{Relevant Literature}

The valuation of non-fungible tokens (NFTs) presents a unique challenge, situated at the intersection of traditional art markets, cryptocurrency economics, and behavioral finance. As unique digital assets, their prices are determined by a complex interplay of factors, including their intrinsic visual characteristics, collection-level attributes such as rarity, and broader market-wide forces. A growing body of literature seeks to disentangle these drivers, providing a foundation for hedonic models that can systematically relate an NFT's observable traits to its market price. This review synthesizes key findings across several relevant domains: the pricing of aesthetic and visual features, the methodological extraction of such features, the influence of collection-level scarcity and brand effects, and the critical role of dynamic market conditions and time-varying premia.

A foundational premise in art valuation is that aesthetic properties are systematically priced. This principle, long established in traditional art markets, provides a crucial starting point for understanding NFT pricing. For instance, research on the works of Jean-Michel Basquiat demonstrates that core color mechanics, such as intensity, luminosity, and contrast, are reliably rewarded by the market, with these premia amplified during the artist’s peak creative periods (Garay et al., 2022a). Similarly, studies of Picasso and Color Field Abstract Expressionists find that specific color palettes and greater color diversity command significant price premiums, confirming that compositional color choices translate directly into economic value (Stepanova, 2019). Recent scholarship confirms that this phenomenon extends directly into the digital realm of NFTs. An analysis of the CryptoPunks collection reveals that specific aesthetic signals—namely, higher color counts and greater textural complexity—are associated with higher prices, while increased brightness and saturation are discounted. These effects persist even after controlling for market conditions and rarity, indicating that visual traits are economically meaningful drivers of value in their own right (Alsultan et al., 2024). However, the relationship is not always straightforward. In studies of abstract paintings, color diversity and contrast are shown to be positively associated with price, whereas generic measures of visual complexity are often insignificant or even negatively correlated (Borisov et al., 2022). This suggests that interpretable compositional features carry more weight than abstract complexity. Furthermore, the explanatory power of handcrafted aesthetic features can be modest and highly dependent on the specific collection being analyzed, underscoring the need for robust models that can account for such heterogeneity (Chen et al., 2023).

Given the importance of visual traits, a key methodological challenge is how to operationalize and measure them effectively. Traditional approaches relying on handcrafted features have been complemented by more advanced computational techniques drawn from computer vision and machine learning. A validated pipeline involves using pre-trained convolutional neural networks (CNNs) to extract high-level visual embeddings from images, which are then compressed using methods like Principal Component Analysis (PCA). This process creates a compact and orthogonal set of features that can be combined with hand-engineered traits and platform signals to improve price prediction (Pala \& Sefer, 2024). This hybrid approach balances the interpretability of classic metrics with the expressive power of deep learning representations. Further refining this, some methodologies fine-tune CNNs to predict human aesthetic ratings directly, yielding model-ready features that align closely with human perceptual judgments and can serve as powerful controls for subjective quality (Talebi \& Milanfar, 2018).

Beyond the aesthetics of an individual token, collection-level attributes play a decisive role in NFT valuation, most notably the concept of trait-based rarity. The distribution of traits within generative art collections is typically highly skewed, meaning genuinely rare items are few. Empirical evidence from millions of transactions shows that these rarer tokens command markedly higher prices, trade less frequently, and generate higher median returns upon resale (Mekacher et al., 2022). In many popular PFP (profile picture) collections, rarity and series-level adoption, measured by the number of unique owners, have been identified as the primary drivers of price. In these contexts, aesthetic signals may be of secondary importance, particularly in the market’s high-price strata (Cui et al., 2023). This highlights the necessity of treating rarity and collection popularity as core control variables in any hedonic model. Failure to do so risks incorrectly attributing the premium associated with scarcity to correlated visual features, thereby biasing the estimated effects of aesthetic traits.

Finally, NFT prices are not determined in a vacuum but are deeply embedded within the volatile cryptocurrency ecosystem and subject to shifting market regimes. The NFT market is significantly influenced by price movements in major cryptocurrencies like Bitcoin and Ethereum, with shocks in these larger markets spilling over to affect NFT sales volume and user activity (Ante, 2022). The market itself exhibits a highly skewed structure, with supply, trading activity, and ownership all following power-law patterns and undergoing clear regime breaks between boom and bust periods (Tang et al., 2023). This inherent instability suggests that the relationships between asset characteristics and price, while significant on average, may not be constant over time. A key conceptual challenge is to separate the effects of a collection’s intrinsic quality from short-run popularity momentum (Park et al., 2024). Static hedonic models serve as a powerful tool for identifying the average, time-invariant premia associated with visual traits and other characteristics. To explore the stability of these premia, dynamic models offer a complementary perspective. For example, a Bayesian dynamic framework applied to the Pop Art market revealed that attribute valuation is state-dependent, changing significantly during the 2008--09 financial crisis (Garay et al., 2022b). This insight is highly relevant for the NFT market, where analysis shows that the primary drivers of valuation can shift between bull, bear, and neutral market phases (Kang \& Lee, 2025). By first establishing baseline effects with a static specification and then exploring their evolution with a dynamic model, a more comprehensive understanding can be achieved, capturing both the average pricing of traits and the state-contingent nature of their valuation.

\section{Data}

\subsection{NFT Collections and Transaction Data}
26 Ethereum NFT collections were drawn from OpenSea's Top 30 at the time of sampling. Each is a large, generative art collection with roughly \(10{,}000\) tokens (some more, some fewer). To keep the dataset tractable yet representative, a token-level random sample was taken within every collection: a 10\% random draw plus an additional 200 randomly selected tokens per collection. The extra 200 serve as a buffer for items later dropped due to missing metadata or other filters. Sampling used a fixed random seed to ensure reproducibility.

The study window spans January 2021--March 2025. Within this window, \textbf{94{,}039} sales covering \textbf{24{,}148 unique NFTs} across the 26 collections are observed after data cleaning. The panel is unbalanced because many tokens never trade and some trade multiple times; each sale is treated as a separate observation. All prices were originally quoted in ETH and converted to USD using the trade-day ETH/USD rate.

\subsection{Image Feature Extraction}

Each image is encoded as a 196--dimensional vector comprising three complementary feature families: 31 classic, interpretable cues; 15 texture descriptors; and 150 compact deep representations. The aim is to summarize visible design choices in a way that is both interpretable and model–ready; brief “for example” explanations follow each technical item.

\noindent\textbf{Classic features (31 total).}
Concretely, the classic block covers:
(i) \emph{color statistics in HLS/HSV} (8), including global hue, lightness, and saturation with diversity diagnostics such as hue range, number of distinct hue bins, and the modal hue with its share. \emph{For example:} a bright neon background exhibits high saturation; a washed-out pastel exhibits low saturation; a predominantly golden figure has a hue near yellow/orange.
(ii) \emph{composition–focus ratios} (3) comparing the central third of the canvas to the full image for hue, lightness, and saturation (\texttt{COMPOSITION\_FOCUS\_*}); values $>1$ indicate concentrated focal design, while $<1$ indicates diffuse layout. \emph{For example:} a centered character on a plain background yields a ratio $>1$, whereas a uniform gradient yields $\approx 1$.
(iii) \emph{edge geometry and strength} (5) via Canny coverage, horizontal/vertical span (\texttt{EDGE\_RANGE\_X/Y}), bounding-box share, and a Sobel energy ratio (edge salience relative to the image mean). \emph{For example:} thick inked outlines raise edge area and energy; soft airbrushed shading lowers both.
(iv) \emph{palette structure from a 6–color K–means} (6), summarized by HSV saturation/value moments, Lab "colorfulness," and mean $\Delta E$ separation among palette centroids. \emph{For example:} cohesive pastels show small $\Delta E$; clashing neons show large $\Delta E$.

\begin{figure}[htbp]
  \centering
  \includegraphics[width=\linewidth]{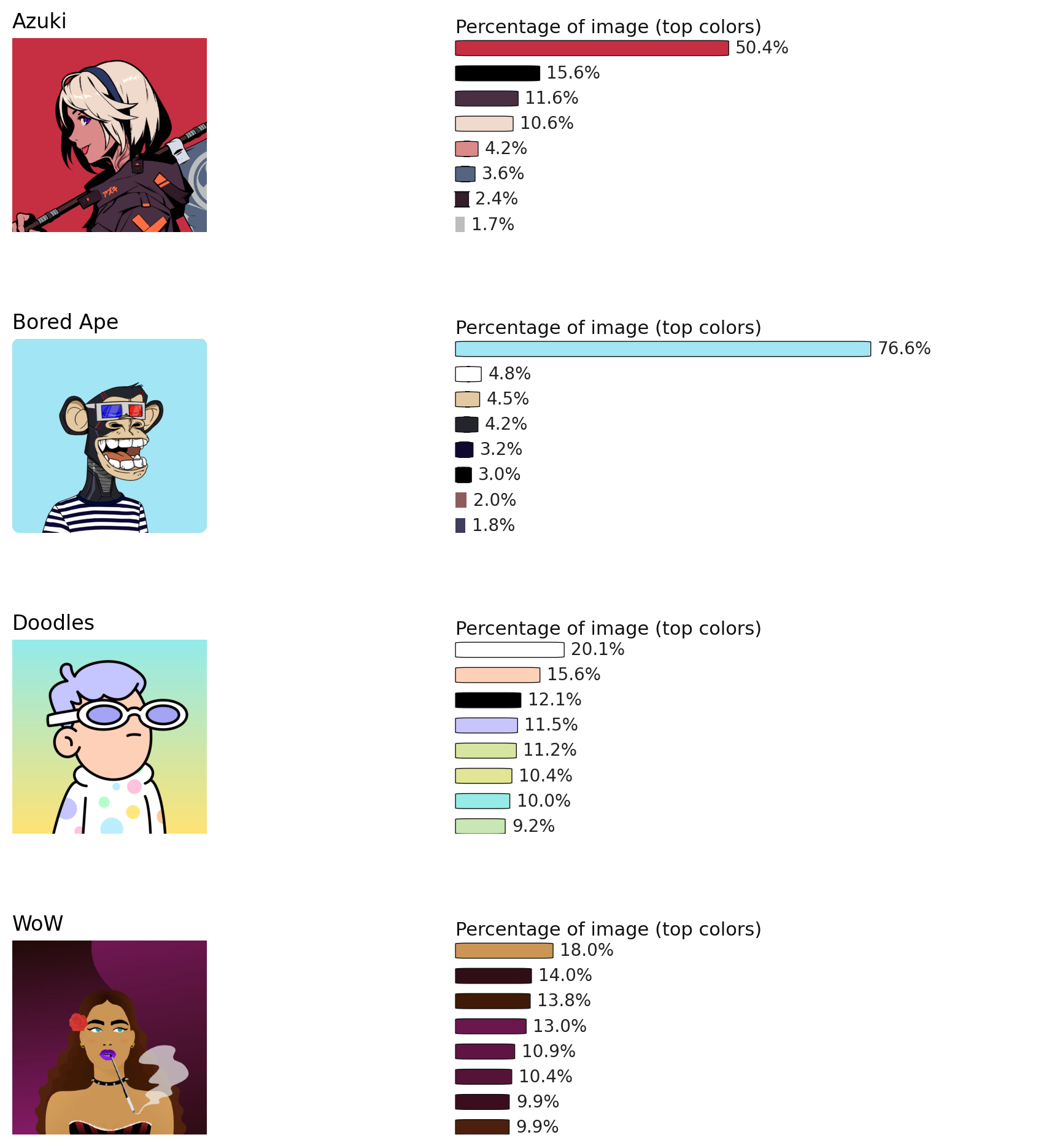}
  \caption{Representative color–palette profiles by collection. Bars show the share of image area captured by each of the top palette colors for a sample image from Azuki, Bored Ape, Doodles, and WoW. This illustrates cross-collection differences in dominant hues and motivates the palette/composition features used in the regressions.}
  \label{fig:palette-panels}
\end{figure}
(v) \emph{line-art geometry} (6) using skeletonization and Hough segments to proxy average stroke length and thickness, straight vs.\ curved prevalence, a box-counting fractal dimension, and endpoints per segment. \emph{For example:} ruler-straight hatching increases straight-segment prevalence; organic drawings increase curved structure.
(vi) \emph{single-number “quality” scalars} (3): Laplacian variance (sharpness), Shannon entropy (tonal richness), and histogram energy (concentration). \emph{For example:} crisp, high-detail drawings have higher sharpness; nearly flat backgrounds have lower entropy but higher energy (mass concentrated in a few gray levels).

\noindent\textbf{Texture features (15 total).}
Two complementary texture families are used:
\emph{Local Binary Patterns} (10-bin histogram) summarize micro-textures; and \emph{FFT ring energies} (5 concentric frequency bands) split spectral power from coarse to fine detail. \emph{For example:} a polka-dot hoodie raises specific LBP bins; fine stippling increases high-frequency FFT bands, whereas broad flat fills increase low-frequency bands.

\noindent\textbf{Deep representations (150 total).}
To capture global content and style, deep representations are included: a 2048–D ResNet descriptor reduced to 100 principal components, and Gram–matrix style features from VGG layers reduced to 50 components. The resulting inventory totals 196 candidate features per image prior to modeling.\footnote{Implementation details include batched inference for CNNs, ImageNet normalization, and consistent pre–processing (resize to 256, center–crop 224). A compact, collection–agnostic script outputs a single CSV (one row per image) with all classic, texture, and deep/PCA features.}

To summarize, classic features align with recognizable design choices (focal saturation, edge salience, curvature, palette cohesion), allowing coefficients to be read as premia/discounts on these attributes, while deep PCs act as orthogonal controls that absorb remaining visual regularities without over-parameterizing the regression.

Two practical choices are worth noting. First, hue is \emph{circular}; whenever hue enters a regression it is encoded via $\sin/\cos$ and the raw angle is omitted to avoid discontinuity and collinearity. Second, no dataset-level standardization is performed inside the feature script; for econometric stability, all numeric regressors are z-scored after merging with returns/sentiment data in the analysis pipeline.

Not all 196 features enter the final regressions. To safeguard parsimony and out-of-sample stability, a small, pre-registered reduction step screens highly collinear candidates and retains a compact subset for the static and dynamic hedonic models. Extracting a hundred-plus descriptors per image across tens of thousands of tokens enables richer questions about how visible design choices and style map into prices---without overfitting the noise.

\subsection{Market Controls}

Control variables are constructed to absorb broad market movements that could confound image–feature effects. Cryptocurrencies such as Ethereum (ETH), Bitcoin (BTC), and Solana (SOL) prices are retrieved from Yahoo Finance and transformed into returns on their close prices, to capture crypto-wide wealth, liquidity, and opportunity-cost channels that co-move with NFT pricing.

Equity-market conditions are proxied by S\&P\,500 and Nasdaq Composite returns (FRED). While correlated, the two indices load differently on broad-market versus tech/growth risk; including both reduces sensitivity to the choice of a single equity benchmark. Expressing these series in returns and standardizing them mitigates shared trends; any remaining collinearity primarily inflates standard errors rather than inducing bias, while still soaking up risk-on/risk-off shocks.

Market sentiment is captured by the Fear \& Greed Index (0–100; 0 = extreme fear, 100 = extreme greed), aligned to the analysis frequency and standardized  (Alternative.me, 2025). All control variables enter the regressions as standardized series.

\subsection{Market Cycles}

NFT trading did not evolve smoothly over the sample; it moved in recognizable phases. To reflect this, the timeline is partitioned into ten contiguous cycles that cover the full period without gaps. In broad terms, the sequence includes an early-2021 run-up, a late-2021/early-2022 peak with initial cooling, the 2022 downturn that spans the Terra/UST and FTX episodes, a brief liquidity burst around the Blur launch (February–March 2023) followed by retrenchment, a late-2023 to early-2024 thaw, a mid-2024 pause, a Q4-2024 rebound, and an early-2025 soft patch. Exact start and end dates are listed in the cycle-interval table. The cycle index is used in two ways: as level controls for regime shifts and as the indexing variable that allows selected coefficients in the dynamic hedonic specification to vary by regime, providing flexibility without month-by-month over-parameterization.

\section{Methodology}

\subsection{Three-Stage Image–Feature Reduction}

The objective is to compress a heterogeneous set of image descriptors into a compact, time-robust subset for downstream hedonic models, while preventing look-ahead and preserving family diversity.

\textbf{Universe and families.} Image features (196 total) are partitioned into families—handcrafted (color, composition, edges, palette, texture), CNN–PCA, and style–Gram–PCA—via name prefixes. Families are used for pruning and representation quotas in later steps. Targets use \(y=\log(1+\text{price})\).

\paragraph{Stage 1: Preliminary screening (variance \& redundancy)}
(1) \emph{Variance threshold}: drop near-constants (variance \(<10^{-4}\)). 
(2) \emph{Family-wise correlation prune}: within each family, if two features have \(|r|>0.95\), retain the one with larger \(|\mathrm{corr}(y,\cdot)|\) and discard the other. This prevents a single large family from crowding out the rest purely by correlation structure.

\paragraph{Stage 2: Time-aware stability via LassoCV}
Observations are sorted by date and split into repeated \emph{contiguous} time blocks. Within each block, a TimeSeriesSplit (\(K=5\)) selects the L1 penalty, a Lasso is fit on the earlier portion, and a selection event is recorded for every feature with a nonzero coefficient. Repeating this over \(N_{\text{runs}}\) blocks yields a stability score
\[
\text{stability}_j=\frac{1}{N_{\text{runs}}}\sum_{r=1}^{N_{\text{runs}}}\mathbf{1}\{\hat\beta_{j}^{(r)}\neq 0\},
\]
with the mean absolute coefficient (when selected) kept as an auxiliary signal. Standardization is applied only for Lasso fitting; missing values are median-imputed.

\paragraph{Stage 3: One non-linear check (permutation importance)}
A single Random Forest (on Stage-1 features) is trained on a fixed time block, and \emph{permutation importance} is computed on its validation slice (multiple repeats). A guardrail is formed via a lower 95\% bound,
\(\mathrm{PI}^{\mathrm{low}}=\overline{\mathrm{PI}}-1.96\,\mathrm{sd}(\mathrm{PI})\).
Permutation scores are min–max normalized and blended with stability:
\[
\text{score}_j = 0.60\,\text{stability}_j + 0.40\,\widetilde{\mathrm{PI}}_j.
\]

\paragraph{Final gate and family minima}
High-confidence candidates satisfy \(\text{stability}\ge 0.60\) and \(\mathrm{PI}^{\mathrm{low}}>0\). Soft family minima (e.g., handcrafted\(\ge 4\), CNN\(\ge 6\), style\(\ge 3\)) are enforced before capping the total (up to 20). If needed, thresholds are relaxed in predefined steps to meet minima, and the top blended scores fill remaining slots.

Time-blocked resampling and CV avoid leakage; L1 yields sparse, reproducible selections; a single non-linear check adds a distinct signal at modest cost; family rules preserve interpretability and balance.

\subsection{Static Hedonic Mixed-Effects Model}\label{subsec:static}

\subsubsection{Outcome and Core Specification}
Let $i$ index NFTs, $c(i)$ their collection, and $t$ the transaction date. The dependent variable is the log--price transform
\[
y_{it}=\log\!\bigl(1+\text{price}_{it}\bigr),
\]
which stabilizes variance and accommodates strictly positive prices while retaining observations at very low levels.

The baseline hedonic specification decomposes $y_{it}$ into observed controls and image features, month fixed effects, and two crossed random intercepts:
\[
y_{it}
= \alpha 
+ \mathbf{x}^{\top}_{it}\boldsymbol{\beta} 
+ \delta_{m(t)} 
+ u_{i} 
+ v_{c(i)} 
+ \varepsilon_{it},
\]
\noindent where $\mathbf{x}_{it}$ stacks the returns-only controls (ETH, BTC, SOL, S\&P 500, NASDAQ returns, and the Fear \& Greed index) together with the selected image features retained by the upstream selection pipeline; $\delta_{m(t)}$ denotes calendar-month fixed effects (factor $C(\text{month})$); $u_i$ is an NFT-level random intercept capturing persistent token-specific price levels; $v_{c(i)}$ is a collection-level random intercept capturing systematic differences across collections; and $\varepsilon_{it}$ is idiosyncratic error.

Distributional assumptions follow a standard linear mixed-effects structure:
\[
u_i \stackrel{\text{i.i.d.}}{\sim} \mathcal{N}(0,\sigma_{\text{nft}}^2),\qquad
v_{c} \stackrel{\text{i.i.d.}}{\sim} \mathcal{N}(0,\sigma_{\text{coll}}^2),\qquad
\varepsilon_{it} \stackrel{\text{i.i.d.}}{\sim} \mathcal{N}(0,\sigma^2),
\]
with $u_i$, $v_c$, and $\varepsilon_{it}$ mutually independent. This specification partially pools information within NFTs and within collections while controlling flexibly for common month-specific shocks.

\subsubsection{Feature Engineering and Design Matrix}

\noindent Circular color variable such as- angle-like hue measures (e.g., \texttt{COLOR\_MOST\_FREQUENT\_HUE}) are encoded via sine and cosine,
\[
h_{it}^{(\sin)}=\sin(\theta_{it}),\qquad h_{it}^{(\cos)}=\cos(\theta_{it}),
\]
where $\theta_{it}$ is interpreted in radians after unit detection (degrees wrapped to $[0,360)$ then converted; otherwise treated as cycles in $[0,1]$). This avoids artificial discontinuities at the $0/2\pi$ boundary and permits linear modeling of circular information; the raw angle is \emph{excluded} from $\mathbf{x}_{it}$ to prevent collinearity with its $\sin/\cos$ pair. 
\paragraph{}All numeric regressors in $\mathbf{x}_{it}$ are \emph{z-scored} (mean-zero, unit-variance) prior to estimation, so coefficients $\beta_k$ describe the marginal effect of a one–standard-deviation change in regressor $k$ on log-price, conditional on fixed and random effects; standardization also improves numerical conditioning of the mixed-effects optimizer. Potential multicollinearity is screened by flagging regressor pairs with $|\rho|>0.95$; if the full model fails to converge, a pre-specified simplification drops the tightest index pair (S\&P~500, NASDAQ) and limits high-dimensional image blocks, preserving identifiability without changing the estimand.

\subsubsection{Fixed and Random Effects Implementation}

\noindent Fixed calendar effects are included by specifying $y \sim \mathbf{x} + C(\text{month})$, where $C(\cdot)$ denotes a categorical factor. This controls for seasonality and market-wide shifts at a monthly frequency without imposing parametric dynamics.

\noindent Random effects follow a crossed structure: an NFT-level random intercept implemented with \texttt{groups = nft\_id}, and a collection variance component. The latter yields independent collection-specific intercepts that share a common variance parameter $\sigma_{\text{coll}}^2$. Taken together, $(u_i, v_{c(i)})$ absorb persistent heterogeneity at the token and collection levels.

\noindent Observations satisfy positive price, non-missing identifiers (\texttt{nft\_id}, \texttt{collection\_code}) and date, and complete information for the dependent variable, month index, group indicators, and all regressors required by the formula. Rows with $\pm\infty$ or NaN in any of these fields are dropped. The month index is computed from the transaction date at calendar-month granularity.

\noindent Because regressors are standardized, a coefficient $\beta_k$ is the \emph{semi-elasticity} of log-price with respect to a one–SD change in predictor $k$. Month fixed effects $\delta_{m}$ capture common shifts in pricing levels unrelated to features or returns. The NFT and collection random intercepts remove persistent level differences at both granularities, mitigating omitted-variable bias from unobserved, time-invariant attributes.

\subsection{Bayesian Dynamic Hedonic Mixed-Effects Model}

Let $i$ index NFTs, $c(i)$ their collection, and $t$ the transaction date. Let $\tau=\tau(t)\in\{1,\dots,T\}$ denote the \emph{market cycle} in which $t$ falls (defined by exogenous, hand-labeled calendar breaks covering January~2021 through March~2025).\footnote{The cycle partition is fixed prior to estimation and used only to index time effects and time-varying parameters. In the empirical implementation there are $T=10$ cycles; labels are for exposition only and play no role in identification.}
The dependent variable is the log transform
\[
y_{it}=\log\!\bigl(1+\text{price}_{it}\bigr),
\]
as in the static model.

Regressors were partitioned into two blocks: (i) a \emph{static} block $\mathbf{x}_{it}\in\mathbb{R}^{P}$ of standardized controls (market returns, sentiment) and selected image features whose coefficients are constant across cycles; and (ii) a \emph{TVP} block $\mathbf{z}_{it}\in\mathbb{R}^{K}$ (e.g., \emph{Composition Focus Saturation}) whose coefficients are allowed to vary by cycle. All numeric regressors are z-scored prior to estimation.

\subsubsection{Core Equation}
The observation equation augments the static hedonic with cycle effects and time-varying coefficients:
\begin{equation}
\label{eq:dyn-hedonic}
y_{it}
= \alpha
+ \mathbf{x}_{it}^{\top}\boldsymbol{\gamma}
+ \mathbf{z}_{it}^{\top}\boldsymbol{\beta}_{\tau(t)}
+ \delta_{\tau(t)}
+ u_{c(i)}
+ w_{c(i),\,\tau(t)}
+ \varepsilon_{it},
\end{equation}
\noindent Here, $\alpha$ is an intercept; $\boldsymbol{\gamma}\in\mathbb{R}^{P}$ are \emph{static} coefficients; $\boldsymbol{\beta}_{\tau}\in\mathbb{R}^{K}$ are \emph{cycle-specific} (TVP) coefficients; $\delta_{\tau}$ are \emph{cycle effects} that capture level shifts common to all NFTs in cycle $\tau$, subject to a sum-to-zero constraint for identification; $u_{c}$ is a \emph{collection} random intercept absorbing persistent brand; $w_{c,\tau}$ is an \emph{optional} collection$\times$cycle deviation (collection-specific drift over cycles); and $\varepsilon_{it}$ is idiosyncratic noise.

\subsubsection{Priors and State Evolution}

\paragraph{Cycle effects.}
We impose centered cycle effects with partial pooling:
\[
\delta_{\tau}=\sigma_{\text{cycle}}\Bigl(\delta^{(0)}_{\tau}-\tfrac{1}{T}\sum_{s=1}^{T}\delta^{(0)}_{s}\Bigr),
\qquad
\delta^{(0)}_{\tau}\stackrel{\text{i.i.d.}}{\sim}\mathcal{N}(0,1),
\qquad
\sigma_{\text{cycle}}\sim\mathcal{H}\mathcal{N}(0.5).
\]

\paragraph{Static coefficients.}
\[
\gamma_{p}\stackrel{\text{i.i.d.}}{\sim}\mathcal{N}(0,\,0.5^2),\qquad p=1,\dots,P.
\]

\paragraph{Time-varying coefficients (TVP).}
For each TVP regressor $k\in\{1,\dots,K\}$, the coefficient evolves across cycles as a first-order Gaussian random walk around a pooled mean:
\begin{align}
\bar\beta_k &\sim \mathcal{N}(0,\,0.5^2), \quad
\sigma_{\bar{k}} \sim \mathcal{H}\mathcal{N}(0.1), \quad
\beta_{k,1} \sim \mathcal{N}(\bar\beta_k,\,\sigma_{\bar{k}}^2), \notag \\
\omega_k &\sim \mathcal{H}\mathcal{N}(0.07), \notag \\
\beta_{k,\tau} \mid \beta_{k,\tau-1},\omega_k &= \beta_{k,\tau-1} + \varepsilon_{k,\tau}, \quad
\varepsilon_{k,\tau}\stackrel{\text{i.i.d.}}{\sim}\mathcal{N}(0,\,\omega_k^2), \quad \tau=2,\dots,T. \notag
\end{align}
Collecting $\boldsymbol{\beta}_{\tau}=(\beta_{1,\tau},\dots,\beta_{K,\tau})^{\top}$ yields the $K\times T$ state matrix $\mathbf{B}$ that governs $\mathbf{z}_{it}^{\top}\boldsymbol{\beta}_{\tau(t)}$ in \eqref{eq:dyn-hedonic}. The hyperparameters $(\bar\beta_k,\sigma_{\bar{k}},\omega_k)$ jointly control the across-cycle average level, the dispersion around that average, and the smoothness (cycle-to-cycle volatility) of the path.

\paragraph{Random intercepts}
\[
u_{c}\stackrel{\text{i.i.d.}}{\sim}\mathcal{N}(0,\,\sigma_{\text{coll}}^2),\qquad
\sigma_{\text{coll}}\sim\mathcal{H}\mathcal{N}(0.5).
\]
NFT-level random intercepts $u_i\sim\mathcal{N}(0,\sigma_{\text{nft}}^2)$ can be added when repeated trades per NFT are frequent.

\paragraph{Collection$\times$cycle deviation (brand drift)}
To allow brand preferences to drift by cycle, the following has been included,
\begin{align}
w_{c,\tau} &= \sigma_{\text{coll}\times\text{cycle}}\,
\Bigl(v_{c,\tau}-\tfrac{1}{T}\sum_{s=1}^{T} v_{c,s}\Bigr), \notag \\
v_{c,\tau} &\stackrel{\text{i.i.d.}}{\sim}\mathcal{N}(0,1), \quad
\sigma_{\text{coll}\times\text{cycle}} \sim\mathcal{H}\mathcal{N}(0.2). \notag
\end{align}
which centers each collection’s cycle profile to prevent confounding with $u_c$ and $\delta_{\tau}$. 

\paragraph{}
Heavy-tailed measurement noise has been used to reduce sensitivity to outliers:
\begin{align}
\varepsilon_{it}\,\big|\,\nu,\sigma &\sim t_{\nu}(0,\sigma), \quad
\sigma\sim\mathcal{H}\mathcal{N}(1), \quad
\nu=2+\eta, \quad \eta\sim\text{Exp}(1/28). \notag
\end{align}

\subsubsection{Identification, interpretation, and design choices}
\noindent The cycle effects $\{\delta_{\tau}\}$ are sum-to-zero centered for level identification, and the TVP states $\{\boldsymbol{\beta}_{\tau}\}$ are identified through their random-walk evolution and associated priors. With standardized regressors, a TVP coefficient $\beta_{k,\tau}$ is the semi-elasticity of log-price with respect to a one–SD change in regressor $k$ \emph{within cycle $\tau$}; the static coefficients $\boldsymbol{\gamma}$ carry the analogous interpretation averaged across cycles. Random effects $u_c$ absorb persistent collection-level differences, and the optional $w_{c,\tau}$ term captures collection-specific deviations around the common cycle effects, i.e., cycle-dependent brand drift.

\noindent Feature handling mirrors the static specification: circular hue variables are encoded as $\sin/\cos$ pairs with the raw angle excluded to avoid collinearity, and all regressors are z-scored using the analysis sample. To reduce $N$ without changing identification, repeated trades within an NFT$\times$cycle cell are aggregated to a single row (median of $y_{it}$ and means of regressors). Multicollinearity is monitored; if numerical issues arise, highly correlated market indices (e.g., S\&P~500 and NASDAQ) can be pruned ex ante without altering the estimand.

\noindent Allowing $\boldsymbol{\beta}_{\tau}$ to evolve across cycles lets the price impact of salient visual traits (such as composition focus/saturation) strengthen or weaken with market regimes rather than being forced into a single average effect. The optional collection$\times$cycle component $w_{c,\tau}$ separates brand-specific cycle drift from the global cycle effects $\delta_{\tau}$, mitigating the risk of attributing brand dynamics to either features or the common time effect.

\section{Results}

\subsection{Static Hedonic Mixed Effect Model}

The study asks whether economically interpretable \emph{visual characteristics} and \emph{market conditions} are systematically related to NFT prices after accounting for time and hierarchical structure. The static mixed–effects estimates provide a clear answer: (i) risk–sentiment and selected crypto returns co-move with realized prices within the month; (ii) several transparent image traits—salience, curvature, focal saturation, and palette cohesion—are strongly associated with higher prices, while visual clutter, thick line work, and dispersed palettes are discounted; and (iii) a substantial share of price variation reflects persistent heterogeneity at the NFT and collection levels, justifying the hierarchical design.

\begin{table}[htbp]
\centering
\small
\begin{threeparttable}
\caption{Static Hedonic Mixed-Effects Model (key coefficients; month fixed effects omitted)}
\label{tab:static-mixedlm}
\begin{tabular}{
    >{\raggedright\arraybackslash}p{0.45\textwidth}
    S[table-format=+1.3]
    S[table-format=1.3]
    S[table-format=+2.3]
    S[table-format=1.3]
    S[table-format=+1.3]
    S[table-format=+1.3]
}
\toprule
\multicolumn{1}{c}{Variable} &
\multicolumn{1}{c}{Coef.} &
\multicolumn{1}{c}{SE} &
\multicolumn{1}{c}{$z$} &
\multicolumn{1}{c}{$p$} &
\multicolumn{1}{c}{CI Low} &
\multicolumn{1}{c}{CI High} \\
\midrule
Intercept & 4.663 & 0.519 & 8.978 & 0.000 & 3.645 & 5.681 \\
\addlinespace[2pt]
\multicolumn{7}{l}{\textit{Market covariates}}\\
ETH\_return            & -0.033 & 0.009 & -3.717 & 0.000 & -0.050 & -0.016 \\
BTC\_return            & -0.026 & 0.008 & -3.097 & 0.002 & -0.043 & -0.010 \\
SOL\_return            &  0.060 & 0.006 &  9.841 & 0.000 &  0.048 &  0.072 \\
SP500\_return          &  0.010 & 0.014 &  0.741 & 0.459 & -0.017 &  0.038 \\
NASDAQCOM\_return      & -0.017 & 0.014 & -1.204 & 0.229 & -0.045 &  0.011 \\
fear\_greed\_index     &  0.152 & 0.009 & 16.013 & 0.000 &  0.133 &  0.170 \\
\addlinespace[2pt]
\multicolumn{7}{l}{\textit{Image features (selected)}}\\
\emph{Edge Bounding Box Area}       &  1.013 & 0.083 & 12.256 & 0.000 &  0.851 &  1.175 \\
\emph{Edge Range X}                  & -0.280 & 0.056 & -5.034 & 0.000 & -0.389 & -0.171 \\
\emph{Edge Range Y}                  & -0.078 & 0.038 & -2.033 & 0.042 & -0.153 & -0.003 \\
\emph{Line Art Average Thickness}       & -0.473 & 0.015 & -32.022& 0.000 & -0.501 & -0.444 \\
\emph{Line Art Proportion Curved}         &  0.390 & 0.014 & 27.867 & 0.000 &  0.363 &  0.418 \\
\emph{Composition Focus Saturation}  &  0.119 & 0.009 & 13.437 & 0.000 &  0.101 &  0.136 \\
\emph{Composition Focus Lightness}   & -0.102 & 0.009 & -11.655& 0.000 & -0.119 & -0.085 \\
\emph{Color Lightness Distribution} & -0.201 & 0.009 & -21.908& 0.000 & -0.219 & -0.183 \\
\emph{Color Hue Distribution}      & -0.099 & 0.009 & -11.197& 0.000 & -0.116 & -0.081 \\
\emph{Color Palette Mean Delta-E}  & -0.076 & 0.009 & -8.502 & 0.000 & -0.093 & -0.058 \\
\emph{Color Palette Saturation Std Dev} & -0.047 & 0.010 & -4.554& 0.000 & -0.067 & -0.027 \\
\emph{Color Most Frequent Hue (sin)} & -0.028 & 0.010 & -2.931& 0.003 & -0.047 & -0.009 \\
\emph{Color Most Frequent Hue (cos)} & -0.066 & 0.009 & -7.092& 0.000 & -0.084 & -0.048 \\
\addlinespace[2pt]
\multicolumn{7}{l}{\textit{Deep embeddings (examples)}}\\
\emph{Style Gram PCA Component 2} &  0.006 & 0.008 &  0.779 & 0.436 & -0.010 &  0.022 \\
\emph{CNN PCA Component 34}        & -0.005 & 0.008 & -0.592 & 0.554 & -0.021 &  0.011 \\
\emph{CNN PCA Component 42}        &  0.016 & 0.008 &  1.932 & 0.053 & -0.000 &  0.032 \\
\emph{CNN PCA Component 56}        &  0.002 & 0.008 &  0.268 & 0.789 & -0.014 &  0.018 \\
\emph{CNN PCA Component 57}        &  0.012 & 0.008 &  1.411 & 0.158 & -0.005 &  0.028 \\
\emph{CNN PCA Component 58}        &  0.002 & 0.008 &  0.224 & 0.822 & -0.014 &  0.018 \\
\emph{CNN PCA Component 80}        &  0.006 & 0.008 &  0.751 & 0.452 & -0.010 &  0.022 \\
\bottomrule
\end{tabular}
\begin{tablenotes}[flushleft]
\footnotesize
\item \emph{Notes:} Dependent variable is $y=\log(1+\text{price})$. Observations $N=67{,}072$; NFT groups $=21{,}818$ (min $=1$, max $=91$, mean $=3.1$). Log-likelihood $=-104{,}174.857$; residual variance (scale) $=0.789$; convergence: Yes. \textbf{Month fixed effects are included (baseline 2021--01) but omitted to conserve space.} All continuous regressors are standardized (z-scores). Reported are coefficient estimates, standard errors, $z$-statistics, $p$-values, and 95\% confidence intervals. Random-effects variances: NFT $=0.591$; Collection $=0.591$.
\end{tablenotes}
\end{threeparttable}
\end{table}
\noindent\textit{Overall fit and hierarchical structure.}
The model converges with residual variance (scale) close to $0.79$ and sizeable random-intercept variances at both NFT and collection levels (each $\widehat{\sigma}^2 \approx 0.59$). A simple intra-class correlation (ICC),
\[
\text{ICC} \;=\; \frac{\widehat{\sigma}^2_{\text{NFT}}+\widehat{\sigma}^2_{\text{collection}}}{\widehat{\sigma}^2_{\text{NFT}}+\widehat{\sigma}^2_{\text{collection}}+\widehat{\sigma}^2_\varepsilon}
\;\approx\; 0.60,
\]
implies that roughly sixty percent of the variation in $\log(1+\text{price})$ is attributable to stable NFT- or collection-specific components rather than purely idiosyncratic noise. This supports allowing for both within-collection dispersion and NFT-specific premia (“brand” or provenance effects).

\medskip
\noindent\textit{Standardization and economic magnitudes.} All continuous regressors are standardized; coefficients therefore capture semi-elasticities: for a one–standard deviation increase in a regressor $x$, the implied percent change in $1+\text{price}$ is $\exp(\beta_x)-1$. This places cross-feature comparisons on a common scale.

\medskip
\noindent\textit{Effect sizes.} Because regressors are standardized, a one–SD increase changes expected price by $\exp(\hat\beta)-1$; for example, the \emph{Fear \& Greed index} with $\hat\beta=0.152$ implies $\exp(0.152)-1=0.164\approx\mathbf{16.4\%}$ higher price, while \emph{line–art average thickness} with $\hat\beta=-0.473$ implies $\exp(-0.473)-1=-0.377\approx\mathbf{-37.7\%}$. Two other notable magnitudes are \emph{SOL return} ($\hat\beta=0.060 \Rightarrow \mathbf{+6.2\%}$) and \emph{proportion curved line–art} ($\hat\beta=0.390 \Rightarrow \mathbf{+47.7\%}$) for a one–SD increase.

\medskip
\noindent\textit{Market covariates.} Returns on major crypto assets display an asymmetric pattern: \textit{SOL} return loads positively and with high precision, while \textit{ETH} and \textit{BTC} returns load negatively within the month once time fixed effects are absorbed. Broad equity benchmarks (S\&P~500, NASDAQ) are statistically indistinguishable from zero in this specification. The \textit{Fear \& Greed Index} enters strongly and positively, indicating that risk appetite co-moves with realized NFT prices even after controlling for month effects. Conditional on monthly regimes, crypto-specific conditions and risk sentiment are more informative for NFT pricing than broad equity indices.

\medskip
\noindent\textit{Image features and willingness-to-pay for design.} Several interpretable image traits exhibit large and precisely estimated associations with price. Salience and concentration matter: when edge-defined content occupies a larger fraction of the canvas (\emph{Edge Bounding Box Area}), prices are substantially higher, whereas wider spatial spread of edges along horizontal and vertical axes (\emph{Edge Range X}, \emph{Edge Range Y}) is penalized, pointing to a premium for concentrated, salient composition and a discount for visually diffuse layouts.

Line geometry is priced: thicker line work (\emph{Line Art Average Thickness}) is strongly discounted, while a higher share of curved line segments (\emph{Line Art Proportion Curved}) is rewarded—consistent with willingness-to-pay for curvature over heavy, blocky strokes.

Color and palette cohesion also command premia: greater focal-region saturation (\emph{Composition Focus Saturation}) is rewarded. Conversely, larger spreads in lightness and hue (\emph{Color Lightness Distribution}, \emph{Color Hue Distribution}) are discounted, as is higher palette divergence (\emph{Color Palette Mean Delta-E}, \emph{Color Palette Saturation Standard Deviation}).

The joint significance of the sine and cosine of the dominant hue indicates directional sensitivity on the color wheel; overall, cohesive palettes with saturated focal areas are favored over extreme contrast and dispersion.

\medskip
\noindent\textit{Deep visual embeddings in the presence of explicit features.} Once explicit geometry/color descriptors and month effects are included, most CNN/Gram principal components are statistically weak in the static cross-section. This does not rule out relevance in alternative specifications (e.g., interactions with states or in dynamics), but it indicates limited incremental explanatory power here beyond transparent design primitives.

\medskip
\noindent\textit{Economic interpretation and testable implications.} Two messages emerge. First, sentiment matters within the month: conditional on month effects, risk appetite (Fear \& Greed) and selected crypto returns (especially SOL) co-move with realized prices, whereas broad equity indices do not—implicating conditions proximate to crypto–NFT markets rather than general equities.

Second, design traits carry premia in levels: salience, curvature, focal saturation, and palette cohesion are rewarded, while clutter, thickness, extreme contrast, and dispersion are penalized. If these characteristics primarily capture tastes or convenience yields rather than compensation for risk, portfolios tilted toward high-priced traits should, controlling for standard market exposures, earn lower average subsequent returns, with the converse for discounted traits; if instead they proxy for priced risk exposures, future return patterns should align with their conditional covariances with the stochastic discount factor.

The sizeable random intercepts are consistent with persistent brand premia at the NFT and collection levels; whether these premia are taste- or risk-based is empirical, motivating characteristic-sorted portfolios and conditional tests that relate these premia to subsequent returns and state dependence.

\medskip
\noindent\textit{Bridge to dynamics.} Because month fixed effects absorb regime shifts only coarsely, the time-varying parameter (TVP) analysis in the next section asks whether the \emph{strength} (and sometimes the \emph{sign}) of key associations varies across market cycles. The static results identify which interpretable features and controls matter on average; the dynamic specification then probes whether their pricing power is state contingent, addressing the temporal dimension of the research question.

\subsection{Dynamic hedonic model with time-varying parameters}

To complement the static specification, we estimate a Bayesian hierarchical hedonic model (Appendix A: Table A1) in which a selected image attribute follows a time-varying parameter (TVP) across economically meaningful \emph{NFT market cycles}. The estimation sample aggregates repeated trades to one row per NFT\(\times\)cycle (median \(\log(1+\text{price})\), mean covariates), yielding \(N=37{,}126\) observations across ten cycles and 26 collections. Identification mirrors the static design (standardized regressors; same controls and image features), but replaces month dummies with \emph{cycle} effects constrained to sum to zero, admits heavy-tailed Student-\(t\) errors, and includes hierarchical structure at the collection level with collection and collection\(\times\)cycle random effects. Sampling uses NumPyro/JAX NUTS (200 post-tune draws \(\times\) 2 chains; 200 tune; target accept 0.80).

\paragraph{\normalfont\itshape Why Composition Focus Saturation as TVP.} This variable proxies how concentrated and saturated the focal area of an image is—an attribute plausibly sensitive to regime-dependent tastes and attention. It showed economically meaningful effects in the static screening, yet with signs that could vary by market state; letting its slope drift across cycles tests whether its premium is state-dependent rather than fixed.

\begin{figure}[htbp]
  \centering
  \includegraphics[width=\linewidth]{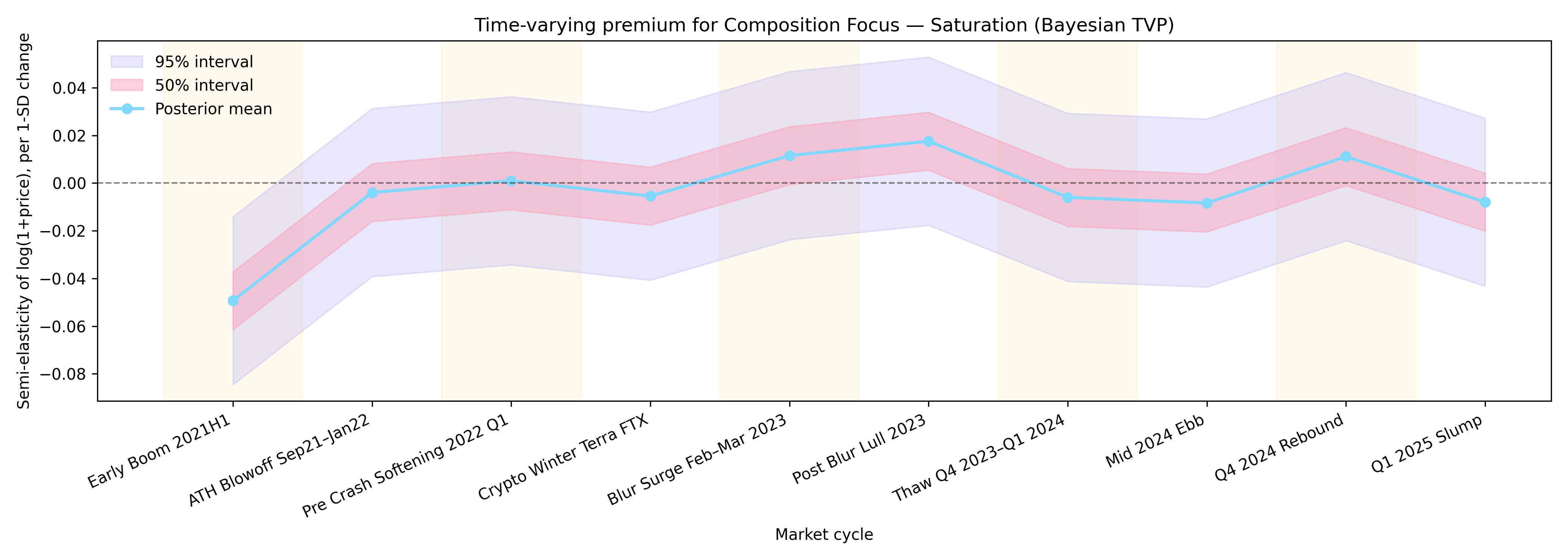}
  \caption{Time-varying premium for \emph{Composition Focus---Saturation} from the Bayesian TVP model. Points are posterior means of the semi-elasticity of $\log(1+\text{price})$ per one--standard-deviation change in the standardized regressor; bands are 50\% and 95\% credible intervals. The path is near zero on average, modestly positive around the \textit{Post Blur Lull 2023} state, and slightly negative in \textit{Mid 2024 Ebb} and \textit{Q1 2025 Slump}, consistent with state-dependent pricing.}
  \label{fig:tvp-composition-focus-saturation}
\end{figure}

The figure tracks how the ``Composition Focus---Saturation'' trait is priced over time. On average, the effect is very small (hovering around zero), but it moves with market mood: in hot phases the trait earns a \emph{small premium}, and in cool phases that premium shrinks or turns into a \emph{small discount}. Specifically, \textit{Early Boom 2021H1} shows a discount (about $-5\%$ per one--standard-deviation increase in the trait); \textit{ATH Blowoff} and \textit{Pre-Crash} are near zero; a modest premium appears around \textit{Post Blur Lull 2023} (peaking near $+1$--$2\%$); \textit{Mid 2024 Ebb} is slightly negative; \textit{Q4 2024 Rebound} shows a small positive blip; and \textit{Q1 2025 Slump} is mildly negative. Most intervals cross zero, so the grand-mean effect is essentially zero. The salient feature is the \emph{sign flip} across regimes, consistent with style premia expanding in risk-on periods and compressing in risk-off periods.

\paragraph{\normalfont\itshape Key coefficients and magnitudes.} (Appendix A: Table A1) Within-state sentiment remains strongly priced: the \emph{Fear \& Greed} coefficient is \(0.116\) (posterior SD \(0.006\)), about \(\exp(0.116)-1\approx 12\%\) higher price for a 1–SD rise in risk appetite. Geometric salience is rewarded: \emph{Edge Bounding Box Area} \(=0.200\) (\(\approx 22\%\)), while spatial dispersion is penalized: \emph{Edge Range X} \(=-0.109\) (\(\approx-10\%\)) and \emph{Edge Range Y} \(=-0.061\) (\(\approx-6\%\)). \emph{Line-Art Prop Curved} shrinks relative to the static model and centers slightly below zero (\(-0.075\), \(\approx-7\%\)), indicating that part of the static curvature premium is re-attributed once state and brand drift are explicit. Among market controls, \emph{SOL return} is modestly negative on average (\(-0.011\), \(\approx-1\%\)), whereas \emph{ETH} and \emph{BTC} returns are close to zero after conditioning on cycles and collection\(\times\)cycle movements.

\paragraph{\normalfont\itshape State dependence (cycles).} Cycle effects quantify level shifts otherwise soaked up by month dummies. The strongest premium occurs in \emph{Pre-Crash Softening 2022Q1} (\(\delta\approx 1.182\), \(\exp(1.182)-1\approx 226\%\)), with \emph{ATH Blowoff Sep21–Jan22} also elevated (\(\approx 59\%\)). Discounts are sizable in \emph{Mid 2024 Ebb} (\(\approx -41\%\)) and \emph{Q1 2025 Slump} (\(\approx -39\%\)). The remaining states are directionally consistent with the narrative but shrink toward the grand mean under aggregation and priors (Appendix A: Table A2).

\paragraph{\normalfont\itshape Brand premium and drift.} The collection random intercept SD is \(\sigma_{\text{coll}}\approx 1.283\), indicating substantial cross-sectional brand premia/discounts. More importantly, the collection\(\times\)cycle SD is \(\sigma_{\text{coll}\times\text{cycle}}\approx 0.912\), implying those brand premia themselves \emph{drift with market state}. This re-attributes variance away from fixed slopes that looked large in the pooled cross-section, explaining why some static image-trait effects attenuate here.

\paragraph{\normalfont\itshape Fit and tails.} The residual SD concentrates around \(0.420\) under Student-\(t\) errors with \(\nu\approx 2.42\), consistent with fat-tailed prices and improved robustness to extremes.

\paragraph{\normalfont\itshape Economic interpretation of the dynamic model.} The TVP specification treats market conditions as evolving regimes and lets the marginal willingness to pay for visual attributes move with those regimes. Because regressors are standardized, slopes read as semi-elasticities; time variation in those slopes indicates that tastes and risk appetite shift across cycles.

\medskip
Two forces organize price formation. First, brand capital is large and not fixed (see \(\sigma_{\text{coll}}\) and \(\sigma_{\text{coll}\times\text{cycle}}\) above), re-allocating variance away from pooled cross-sectional slopes that looked big in static regressions and explaining why some image-trait effects attenuate once state and brand drift are explicit. Second, within-state sentiment is priced (see the Fear \& Greed coefficient above), consistent with risk appetite amplifying willingness-to-pay when liquidity and attention are abundant.

\medskip
Trait premia are conditional rather than universal. The time-varying coefficient on \textit{Composition Focus---Saturation} averages near zero across cycles ($\text{mean}\approx-0.004$, SD $\approx 0.018$), but its sign flips with market conditions: positive in liquidity-rich windows (e.g., Blur surge; the late-2024 rebound) and negative in retrenchment phases (mid-2024 ebb; early-2025 slump). In expansionary, sentiment-rich phases, legible, saturated focal designs are rewarded—matching the idea that attention and speculative demand magnify the payoff to easily processed visual signals—while the same traits are discounted when risk capital steps back and trading thins.

\medskip
Through an asset-pricing lens, the model says prices reflect (i) slow-moving brand heterogeneity, (ii) cycle-specific market tightness, and (iii) trait premia that expand or compress with investor sentiment and trading conditions. For platforms, this implies that curation and homepage real estate should tilt toward high-salience, cohesive designs in hot markets—where those traits command clear premia—and diversify toward subtler, niche aesthetics in cold markets, where the same traits lose pricing power. For portfolio construction, characteristic-sorted strategies should condition on state: long–short tilts toward ``expansion favorites'' (e.g., saturated focal composition) are likely to earn premia only in risk-on regimes, whereas in risk-off regimes the payoff compresses or reverses.

\paragraph{}
In summary, the static model clarifies \emph{which} visual traits and controls are priced on average in the cross-section. The dynamic specification clarifies \emph{when} those premia are high or low and how brand premia evolve with market state: cycle effects capture level shifts, the TVP acts as a trait\(\times\)state interaction, and collection\(\times\)cycle terms quantify brand drift. The two perspectives are complementary: the static model provides interpretable baseline signals; the dynamic model shows those signals are state-contingent and partially mediated by evolving brand premia.

\section{Robustness Checks}

\subsection{Benjamini--Hochberg FDR}

The BH adjustment leaves the paper's substantive inferences intact. Sentiment (\textit{Fear \& Greed}) remains strongly positive, and crypto co-movements matter (positive \textit{SOL} loadings; negative \textit{ETH}/\textit{BTC}). 

\begin{table}[H]
\centering
\caption{Static hedonic mixed-effects with Benjamini--Hochberg FDR (key covariates)}
\label{tab:static_bhfdr}
\begin{threeparttable}
\footnotesize
\begin{tabular}{>{\raggedright\arraybackslash}p{0.45\textwidth}rrrrr}
\toprule
Variable & $\hat\beta$ & SE & $z$ & $p$ & $q_{\mathrm{BH}}$ \\
\midrule
\multicolumn{6}{l}{\textit{Controls}}\\
\addlinespace[2pt]
BTC\_return            & -0.026 & 0.008 & -3.097 & 0.00195 & 0.004$^{***}$ \\
ETH\_return            & -0.033 & 0.009 & -3.717 & 0.00020 & 0.000$^{***}$ \\
SOL\_return            &  0.060 & 0.006 &  9.841 & 0.00000 & 0.000$^{***}$ \\
fear\_greed\_index     &  0.152 & 0.009 & 16.013 & 0.00000 & 0.000$^{***}$ \\
\addlinespace[2pt]
\multicolumn{6}{l}{\textit{Image features}}\\
\addlinespace[2pt]
\emph{Edge Bounding Box Area}        &  1.013 & 0.083 & 12.256 & 0.00000 & 0.000$^{***}$ \\
\emph{Line Art Average Thickness}        & -0.473 & 0.022 & -21.252 & 0.00000 & 0.000$^{***}$ \\
\emph{Line Art Proportion Curved}          &  0.390 & 0.018 & 21.996  & 0.00000 & 0.000$^{***}$ \\
\emph{Composition Focus Saturation}   &  0.119 & 0.009 & 13.100  & 0.00000 & 0.000$^{***}$ \\
\emph{Composition Focus Lightness}    & -0.102 & 0.009 & -11.655 & 0.00000 & 0.000$^{***}$ \\
\emph{Color Lightness Distribution} & -0.201 & 0.009 & -21.908 & 0.00000 & 0.000$^{***}$ \\
\emph{Color Palette Saturation Std Dev}  & -0.046 & 0.010 & -4.554  & 0.00001 & 0.00001$^{***}$ \\
\emph{Color Most Frequent Hue (cos)}  & -0.066 & 0.009 & -7.092  & 0.00000 & 0.00000$^{***}$ \\
\bottomrule
\end{tabular}
\begin{tablenotes}[flushleft]\footnotesize
\item Notes: Dependent variable is $\log(1+\text{price})$. All regressors are standardized. The specification includes month fixed effects and random intercepts for NFT and collection (not shown). $q_{\mathrm{BH}}$ denotes Benjamini--Hochberg adjusted $p$-values within the family of controls and image features. Significance: $^{*}q\le0.10$, $^{**}q\le0.05$, $^{***}q\le0.01$.
\end{tablenotes}
\end{threeparttable}
\end{table}

\medskip
Interpretable visual traits continue to be priced after controlling the false discovery rate: pronounced edge structure and greater curvature load positively; thicker lines, wider lightness dispersion, and higher overall lightness load negatively; composition focus splits by channel (saturation positive, lightness negative). Broad equity benchmarks are not material in this specification (not shown). Overall, multiple-testing control does not overturn the static hedonic results.

\subsection{Dynamic Robustness}
\FloatBarrier
\begin{table}[H]
\centering
\begin{threeparttable}
\caption{Robustness: Cycle-Block Bootstrap for the TVP Average Effect}
\label{tab:tvp_bootstrap}
\begin{tabular}{@{} l c l @{}}
\toprule
\textbf{Measure} & \textbf{Value} & \textbf{Reading} \\
\midrule
Average TVP mean across cycles & $-0.004$ & Near zero on average \\
95\% percentile bootstrap CI\tnote{a} & $[-0.016,\;0.005]$ & CI crosses $0$ (no overall tilt) \\
Share of positive cycles\tnote{b} & $4/10$ & Mixed signs across regimes \\
\bottomrule
\end{tabular}
\begin{tablenotes}[flushleft]
\footnotesize
\item[a] Percentile (nonparametric) bootstrap treating the 10 NFT market cycles as blocks; 50{,}000 resamples.
\item[b] Based on per-cycle posterior means of the time-varying coefficient for \emph{Composition Focus Saturation}. This summarizes direction across cycles; it does not imply per-cycle significance.
\end{tablenotes}
\end{threeparttable}
\end{table}

The cycle-block bootstrap indicates that the \emph{average} time-varying effect of the focal image trait is small and statistically indistinguishable from zero on aggregate (mean $=-0.004$, 95\% CI $[-0.016,\,0.005]$). This is consistent with the dynamic specification, which is designed to allow \emph{cycle-specific} movements rather than to impose a uniform premium. Per-cycle directions are mixed (4 of 10 cycles positive), suggesting that the trait’s pricing relevance varies with market regime. In short, the dynamic hedonic model adds value by capturing \emph{when} the effect matters, even if the grand mean across regimes is near zero.

These checks support a \emph{state-dependent} interpretation: the dynamic structure is warranted to track regime-specific movements, while no non-zero average effect is claimed across all cycles.
\FloatBarrier

\section{Concluding Remarks}

Prices in this market move with both the pixels on the canvas and the phase of the cycle. Across 94{,}039 trades from 26 major Ethereum collections, interpretable visual features—focal saturation, compositional clarity, and smooth curvature—consistently line up with higher valuations, while clutter, heavy line work, and dispersed palettes are discounted. Layering a dynamic, cycle-aware structure on top of the static evidence makes the pattern intuitive: in risk-on windows (e.g., \emph{Blur Surge} and the \emph{Q4 2024 Rebound}), the payoff to legible, high–saturation focal designs expands; in cooler periods (\emph{Mid 2024 Ebb}, \emph{Q1 2025 Slump}), the same traits compress or flip sign. Large collection effects (“brand”) remain, but they drift with state, so part of what looks like a fixed style premium in pooled regressions is better understood as a trait-by-regime effect. The contribution is twofold: a disciplined way to bring high-dimensional, computer-vision features into hedonic models without losing interpretability, and an empirical case that trait premia in digital art are real, economically sized, and \emph{state contingent} rather than universal.

A brief caveat is that the transactions form an unbalanced panel—tokens sell at uneven intervals—so the design absorbs irregular spacing with cycle aggregation, hierarchical structure, and standardized controls; the central conclusions are robust to this feature.

Looking ahead, the framework is a foundation: future work can let regimes be learned rather than labeled, fold in simple microstructure signals when available, and check portability across platforms and categories. These steps would widen scope without changing the main message.

Overall, the evidence supports a clear economic interpretation: observable product characteristics and time-varying market conditions jointly govern price discovery. By unifying transparent image traits with a cycle-aware hedonic design on public ledgers, the study extends tools in cultural economics and asset pricing and offers a practical template for platform strategy and market design in digital goods.

\section*{Acknowledgment}

This project grew at the intersection of two habits: studying economics by day and sketching and painting at night. For a long time, that meant consuming ideas rather than asking new questions. A turning point came when my Ph.D. supervisor, \textbf{Dr.\ Zsolt Becsi}, suggested that a love for artwork could also be a research direction—``pick colors from images, extract features, and build an economic model.'' That simple invitation unlocked the path I follow here. I am also grateful to \textbf{Dr.\ Scott Gilbert}, whose conversations about our shared hobby of art encouraged me to frame the work in a way that speaks to economists. I thank \textbf{Dr.\ Alison Watts} for urging me forward by affirming that the project is interesting. I also thank \textbf{Dr.\ Reza Habib} for providing intuition on NFTs, for steady encouragement, and for helping make the project more intuitive and interesting.

I owe deep thanks to \textbf{Weikang Zhang}, whose encouragement and patient guidance in coding made it possible to collect the data at a large scale, and whose support kept this work moving when the workload felt overwhelming. I am grateful to my brother \textbf{Mirza Sakib} for hands-on help during repeated cleaning phases, and to \textbf{Mirza Samir}, \textbf{Mirza Tariq} (my father), and \textbf{Daulatunnessa Shumi} (my mother) for steady belief and care.

An unexpected nudge also came from outside academia: while posting art on an anonymous Instagram account, I received an attempted NFT-related scam—my first encounter with NFTs. That moment redirected my curiosity toward digital art markets and ultimately shaped the questions asked in this paper.

Any remaining errors are my own. Regardless, I wish to keep learning and growing. To anyone who reads or cites this work—even once—thank you; your attention means the world.

\section*{References}
\begingroup
\setlength{\parindent}{0pt}
\setlength{\parskip}{4pt}
\everypar{\hangindent=0.5in}
\raggedright
\sloppy

Alsultan, S., Kourtis, A., \& Markellos, R. N. (2024). Can we price beauty? Aesthetics and digital art markets. \textit{Economics Letters}, \textit{235}, 111572. \url{https://doi.org/10.1016/j.econlet.2024.111572}

Alternative.me. (n.d.). \textit{Crypto Fear \& Greed Index}. Retrieved September 23, 2025, from https://alternative.me/crypto/fear-and-greed-index/

Ante, L. (2022). The non-fungible token (NFT) market and its relationship with Bitcoin and Ethereum. \textit{FinTech}, \textit{1}(3), 216--224. \url{https://doi.org/10.3390/fintech1030017}

Borisov, M., Kolycheva, V., Semenov, A., \& Grigoriev, D. (2022). The influence of color on prices of abstract paintings. \textit{arXiv preprint} arXiv:2206.04013.

Chen, Y., Ye, Y., \& Zeng, W. (2023). The rich, the poor and the ugly: An aesthetic-perspective assessment of NFT values. In \textit{Proceedings of the 16th International Symposium on Visual Information Communication and Interaction (VINCI 2023)} (pp. 1--8). ACM. \url{https://doi.org/10.1145/3615522.3615545}

Cui, R., Deng, Y., Fu, B., \& Xi, L. (2023). Exploring the price influences of PFP artworks in NFT based on regression analysis. In G. Vilas Bhau et al. (Eds.), \textit{MSEA 2022, ACSR} (Vol. 101, pp. 390--396). \url{https://doi.org/10.2991/978-94-6463-042-8_57}

Dowling, M. (2022a). Fertile LAND: Pricing non-fungible tokens. \textit{Finance Research Letters, 44}, 102096. https://doi.org/10.1016/j.frl.2021.102096

Dowling, M. (2022b). Is non-fungible token pricing driven by cryptocurrencies? \textit{Finance Research Letters, 44}, 102097. https://doi.org/10.1016/j.frl.2021.102097

Dune Analytics. (2025, September). \textit{OpenSea cumulative volume dashboard}. \url{https://www.dune.com/rchen8/opensea}

Entriken, W., Shirley, D., Evans, J., \& Sachs, N. (2018). \textit{ERC-721: Non-Fungible Token Standard (EIP-721)}. Ethereum Improvement Proposals. \url{https://eips.ethereum.org/EIPS/eip-721}

Federal Reserve Bank of St. Louis. (n.d.). \textit{FRED, Federal Reserve Economic Data}. Retrieved September 23, 2025, from https://fred.stlouisfed.org/

Garay, U., P\'erez, E., Casanova, J., \& Kratohvil, M. (2022\textit{a}). Color intensity, luminosity, contrast and art prices: The case of Jean-Michel Basquiat. \textit{Academia Revista Latinoamericana de Administraci\'on}. \url{https://doi.org/10.1108/ARLA-05-2021-0110}

Garay, U., Puggioni, G., Molina, G., \& ter Horst, E. (2022\textit{b}). A Bayesian dynamic hedonic regression model for art prices. \textit{Journal of Business Research}, \textit{151}, 310--323. \url{https://doi.org/10.1016/j.jbusres.2022.06.055}

Kaisto, J., Juutilainen, T., \& Kauranen, J. (2024). Non-fungible tokens, tokenization, and ownership. \textit{Computer Law \& Security Review, 54}, 105996. https://doi.org/10.1016/j.clsr.2024.105996

Kang, H.-J., \& Lee, S.-G. (2025). Market phases and price discovery in NFTs: A deep learning approach to digital asset valuation. \textit{Journal of Theoretical and Applied Electronic Commerce Research}, \textit{20}, 64. \url{https://doi.org/10.3390/jtaer20020064}

Mekacher, A., Bracci, A., Nadini, M., Martino, M., Alessandretti, L., Aiello, L. M., \& Baronchelli, A. (2022). Heterogeneous rarity patterns drive price dynamics in NFT collections. \textit{Scientific Reports}, \textit{12}, 13890. \url{https://doi.org/10.1038/s41598-022-17922-5}

Pala, M., \& Sefer, E. (2024). NFT price and sales characteristics prediction by transfer learning of visual attributes. \textit{The Journal of Finance and Data Science}, \textit{10}, 100148. \url{https://doi.org/10.1016/j.jfds.2024.100148}

Park, J., Lee, Y., Yang, D., Park, J., \& Jung, H. (2024). Artwork pricing model integrating the popularity and ability of artists. \textit{AStA Advances in Statistical Analysis}. \url{https://doi.org/10.1007/s10182-024-00504-3}

Stepanova, E. (2019). The impact of color palettes on the prices of paintings. \textit{Empirical Economics}, \textit{56}, 755--773.

Talebi, H., \& Milanfar, P. (2018). NIMA: Neural image assessment. \textit{IEEE Transactions on Image Processing}, \textit{27}(8), 3998--4011. \url{https://doi.org/10.1109/TIP.2018.2831899}

Tang, M., Feng, X., \& Chen, W. (2023). Exploring the NFT market on Ethereum: A comprehensive analysis and daily volume forecasting. \textit{Connection Science}, \textit{35}(1), 2286912. \url{https://doi.org/10.1080/09540091.2023.2286912}

The Block. (2025). \textit{Ethereum NFT marketplace monthly volume (wash-trading filtered)}. \url{https://www.theblock.co/data/nft-non-fungible-tokens/marketplaces/nft-marketplace-monthly-volume}

Yahoo. (n.d.). \textit{Yahoo Finance}. Retrieved September 23, 2025, from https://finance.yahoo.com/

\endgroup

\clearpage

\section*{Appendix A. Additional Tables}
\addcontentsline{toc}{section}{Appendix A. Additional Tables} 

\setcounter{table}{0}
\renewcommand{\thetable}{A\arabic{table}}

\begin{table}[!htb]
\centering
\footnotesize
\setlength{\tabcolsep}{3pt}
\renewcommand{\arraystretch}{0.85}
\caption{Dynamic Hedonic Price Model with Time-Varying Parameter (NFT cycles)}
\label{tab:dynamic-hedonic}
\begin{threeparttable}
\begin{tabular}{@{}l r@{}}
\toprule
\textbf{Variable} & \textbf{Coefficient (Post.\ SD)} \\
\midrule
\multicolumn{2}{@{}l@{}}{\textit{Dependent variable and sample}}\\
$\log(1+\text{price})$ & \\
Observations: $37{,}126$;\ Collections: $26$;\ Cycles: $10$ & \\
\addlinespace[1pt]
\multicolumn{2}{@{}l@{}}{\textit{Intercept}}\\
Intercept & $7.390^{**}\ (0.287)$ \\
\addlinespace[1pt]
\multicolumn{2}{@{}l@{}}{\textit{Fixed Effects}}\\
ETH Return & $0.006\ (0.006)$ \\
BTC Return & $0.000\ (0.006)$ \\
SOL Return & $-0.011^{**}\ (0.005)$ \\
S\&P~500 Return & $-0.012\ (0.010)$ \\
NASDAQ Return & $0.004\ (0.011)$ \\
Fear \& Greed Index & $0.116^{**}\ (0.006)$ \\
Edge Bounding Box Area & $0.200^{**}\ (0.046)$ \\
Edge Range Y & $-0.061^{**}\ (0.021)$ \\
Edge Range X & $-0.109^{**}\ (0.031)$ \\
Color Hue Distribution & $0.006^{**}\ (0.003)$ \\
Composition Focus Lightness & $-0.000\ (0.003)$ \\
Color Lightness Distribution & $0.003\ (0.004)$ \\
Color Palette Mean $\Delta E$ & $-0.004\ (0.004)$ \\
Line Art Thickness & $0.009\ (0.006)$ \\
Line Art Prop Curved & $-0.075^{**}\ (0.010)$ \\
\emph{Style Gram PCA Component 2} & $0.005\ (0.003)$ \\
\emph{Color Palette Saturation Std Dev} & $-0.001\ (0.003)$ \\
\emph{CNN PCA Component 34} & $-0.004\ (0.003)$ \\
\emph{CNN PCA Component 57} & $-0.000\ (0.003)$ \\
\emph{CNN PCA Component 80} & $0.002\ (0.003)$ \\
\emph{CNN PCA Component 42} & $0.004\ (0.003)$ \\
\emph{CNN PCA Component 56} & $0.000\ (0.003)$ \\
\emph{CNN PCA Component 45} & $-0.001\ (0.003)$ \\
\emph{CNN PCA Component 58} & $0.001\ (0.003)$ \\
\addlinespace[2pt]
\multicolumn{2}{@{}l@{}}{\textit{Time-Varying Parameter (mean across cycles)}}\\
Composition Focus Saturation (TVP mean) & $-0.004\ (0.012)$ \\
\quad TVP SD (across cycles) & $\sigma = 0.018$ \\
\addlinespace[2pt]
\multicolumn{2}{@{}l@{}}{\textit{Variance Components}}\\
Residual SD & $0.420^{**}\ (0.004)$ \\
Collection Random Effects (SD) & $1.283^{**}\ (0.168)$ \\
Collection$\times$Cycle Effects (SD) & $0.912^{**}\ (0.045)$ \\
Student-$t$ Degrees of Freedom $\nu$ & $2.42\ (0.04)$ \\
\addlinespace[2pt]
\multicolumn{2}{@{}l@{}}{\textit{MCMC Settings}}\\
Draws (post-tune), Chains & $200,\ 2$ \\
Tuning draws, Target accept & $200,\ 0.80$ \\
Backend & NumPyro/JAX \\
\bottomrule
\end{tabular}
\begin{tablenotes}\footnotesize
\item Notes: Regressors are standardized; a 1--SD change implies approximately $(\exp(\beta)-1)\times 100\%$ change in expected price. Posterior SDs in parentheses. Cycle dummy coefficients are omitted here for brevity.
\end{tablenotes}
\end{threeparttable}
\end{table}

\begin{table}[!htb]
\centering
\begin{threeparttable}
\caption{Time-varying parameter (TVP) by market cycle for the regressor \textit{Composition Focus—Saturation}. Values are posterior means in log-price units. Positive/negative indicates the sign of the mean effect within each cycle.}
\label{tab:tvp_comp_focus_saturation_cycles}
\small
\begin{tabular}{rlll}
\toprule
 Rank (by mean) &                      cycle & Posterior mean &     Sign \\
\midrule
              1 &        Post\_Blur\_Lull\_2023 &          0.018 & Positive \\
              2 &    Blur\_Surge\_Feb\_Mar\_2023 &          0.012 & Positive \\
              3 &            Q4\_2024\_Rebound &          0.011 & Positive \\
              4 & Pre\_Crash\_Softening\_2022Q1 &          0.001 & Positive \\
              5 &    ATH\_Blowoff\_Sep21\_Jan22 &         -0.004 & Negative \\
              6 &    Crypto\_Winter\_Terra\_FTX &         -0.005 & Negative \\
              7 &       Thaw\_Q4\_2023\_Q1\_2024 &         -0.006 & Negative \\
              8 &              Q1\_2025\_Slump &         -0.008 & Negative \\
              9 &               Mid\_2024\_Ebb &         -0.008 & Negative \\
             10 &          Early\_Boom\_2021H1 &         -0.049 & Negative \\
\bottomrule
\end{tabular}

\begin{tablenotes}[flushleft]
\footnotesize
\item Note: Higher ranks indicate more positive posterior means. HDI/credible intervals are omitted here because the provided file contained malformed interval columns.
\end{tablenotes}
\end{threeparttable}
\end{table}

\begin{longtable}{lc}
  \caption{Time-varying states for Composition Focus Saturation}
  \label{tab:tvp-states-composition-focus-saturation}\\
  \toprule
                       cycle &   mean \\
  \midrule
  \endfirsthead
  \caption[]{Time-varying states for COMPOSITION\_FOCUS\_SATURATION (full listing).} \\
  \toprule
                       cycle &   mean \\
  \midrule
  \endhead
  \midrule
  \multicolumn{2}{r}{{Continued on next page}} \\
  \midrule
  \endfoot
  
  \bottomrule
  \endlastfoot
     ATH\_Blowoff\_Sep21\_Jan22 & -0.004 \\
     Blur\_Surge\_Feb\_Mar\_2023 &  0.012 \\
     Crypto\_Winter\_Terra\_FTX & -0.005 \\
           Early\_Boom\_2021H1 & -0.049 \\
                Mid\_2024\_Ebb & -0.008 \\
         Post\_Blur\_Lull\_2023 &  0.018 \\
  Pre\_Crash\_Softening\_2022Q1 &  0.001 \\
               Q1\_2025\_Slump & -0.008 \\
             Q4\_2024\_Rebound &  0.011 \\
        Thaw\_Q4\_2023\_Q1\_2024 & -0.006 \\
  \end{longtable}

\clearpage
\begin{table}[htbp]
\centering
\small
\caption{Image Feature Categories and Descriptions (Part 1: Classic Features)}
\label{tab:feature-descriptions-1}
\begin{threeparttable}
\renewcommand{\arraystretch}{1.2}
\begin{tabular}{@{}p{0.22\textwidth}p{0.12\textwidth}p{0.58\textwidth}@{}}
\toprule
\textbf{Feature Category} & \textbf{Count} & \textbf{Description and Examples} \\
\midrule
Color in HLS/HSV & 8 & Captures global \textit{hue}, \textit{lightness}, and \textit{saturation} alongside diversity statistics (hue range, number of distinct hue bins, modal hue and its share). \\
& & \textit{Example:} Bright neon background has high saturation; washed-out pastel has low saturation; predominantly golden figure has hue near yellow/orange. \\
\addlinespace[2pt]
Composition focus: center vs.\ whole & 3 & Uses ratios comparing central third of canvas to full image for hue/lightness/saturation (\texttt{COMPOSITION\_FOCUS\_*}). Values $>1$ indicate concentrated focal area, $<1$ indicates diffuse layout. \\
& & \textit{Example:} Centered character on plain background yields ratio $>1$; uniform gradient background yields $\approx 1$. \\
\addlinespace[2pt]
Edges via Canny/Sobel & 5 & Measures fraction of edge pixels and their spatial span (bounding-box share, horizontal/vertical range), together with Sobel energy ratio (edge strength relative to image mean). \\
& & \textit{Example:} Thick inked outlines produce higher edge area and stronger energy; soft airbrushed shading produces lower values. \\
\addlinespace[2pt]
Single-number "quality" scalars & 3 & Includes Laplacian variance (sharpness), Shannon entropy (tonal richness), and histogram energy (concentration) for complementary one-number summaries. \\
& & \textit{Example:} Crisp, high-detail drawing has higher sharpness; nearly flat background has lower entropy but higher energy (most mass in few gray levels). \\
\addlinespace[2pt]
Palette structure with $K{=}6$ colors & 6 & Applies K-means on RGB to isolate six dominant colors; from their HSV/Lab centroids, computes saturation/value moments, Lab "colorfulness," and mean $\Delta E$ separation. \\
& & \textit{Example:} Cohesive pastel set shows small $\Delta E$; mix of clashing neons shows large $\Delta E$. \\
\addlinespace[2pt]
Line-art geometry & 6 & Uses skeletonization and Hough segments to proxy average stroke length and thickness, straight vs.\ curved prevalence, box-counting fractal dimension, and endpoints per segment. \\
& & \textit{Example:} Technical, ruler-straight hatching yields more straight segments; organic drawings yield more curved structure and different fractal score. \\
\addlinespace[3pt]
\multicolumn{3}{@{}l@{}}{\textbf{Classic Features Total:} 31 per image} \\
\bottomrule
\end{tabular}
\begin{tablenotes}[flushleft]
\small
\item \textit{Notes:} All features are extracted from 24{,}148 NFT images across 26 collections. Classic features provide interpretable design choices and are standardized (z-scored) before entering regression models.
\end{tablenotes}
\end{threeparttable}
\end{table}

\begin{table}[htbp]
\centering
\small
\caption{Image Feature Categories and Descriptions (Part 2: Texture and Deep Features)}
\label{tab:feature-descriptions-2}
\begin{threeparttable}
\renewcommand{\arraystretch}{1.2}
\begin{tabular}{@{}p{0.22\textwidth}p{0.12\textwidth}p{0.58\textwidth}@{}}
\toprule
\textbf{Feature Category} & \textbf{Count} & \textbf{Description and Examples} \\
\midrule
Texture: LBP histogram and FFT ring energies & 15 & Employs Local Binary Patterns to summarize micro-textures (10-bin histogram); concentric FFT bands split spectral power from coarse to fine detail. \\
& & \textit{Example:} Polka-dot hoodie pushes specific LBP bins up; fine stippling increases high-frequency FFT bands, whereas broad flat fills increase low-frequency bands. \\
\addlinespace[2pt]
Deep representations for global content and style & 150 & Utilizes pre-trained CNNs with standardized pre-processing (resize to 256, center-crop 224, ImageNet normalization). Global 2048-dimensional descriptor from ResNet-50 (average-pooled convolutional activations) reduced by PCA to 100 components (\textit{CNN\_PCA\_001--100}), summarizing object/shape/layout information; in parallel, VGG-19 Gram-matrix features (second-order co-activations within selected convolutional blocks) reduced to 50 components (\textit{STYLE\_GRAM\_PCA\_001--050}), emphasizing style/texture and color co-occurrence. \\
& & \textit{Example:} Content PC may capture "frontal face vs.\ profile" or "figure vs.\ background layout," while style PC separates "flat vector look" from "brushy/stippled texture." \\
\addlinespace[3pt]
\multicolumn{3}{@{}l@{}}{\textbf{Total Features:} 196 per image (31 classic + 15 texture + 150 deep representations)} \\
\bottomrule
\end{tabular}
\begin{tablenotes}[flushleft]
\small
\item \textit{Notes:} Deep representations capture complex visual patterns that complement the interpretable classic features. All features are standardized (z-scored) before entering regression models.
\end{tablenotes}
\end{threeparttable}
\end{table}

\end{document}